\documentclass[twocolumn,showpacs,preprintnumbers,amsmath,amssymb]{revtex4}

\usepackage{graphicx}
\usepackage{dcolumn}
\usepackage{bm}

\begin{document}


\title{Coulomb blockade of Cooper pair tunneling and parity effects \\ in the Cooper pair transistor}

\author{S. Corlevi$^1$, W. Guichard$^{1,2}$, F.W.J. Hekking$^{2}$, and D.B. Haviland$^1$}
\affiliation{$^1$Nanostructure Physics, Royal Institute of
Technology, 10691 Stockholm, Sweden\\
$^2$University Joseph Fourier and CNRS, B.P. 166, 25 Avenue des
Martyrs, 38042 Grenoble-cedex 09, France\\}

\begin{abstract}
We have measured the Cooper Pair Transistor (CPT) in a tunable
electromagnetic environment consisting of four one-dimensional
SQUID arrays. The transport properties of the CPT in the high
impedance limit, $Z_\mathrm{env}\gg R_{Q}\simeq 6.45~$k$\Omega$,
are studied for different ratios of the Josephson coupling energy
to the charging energy.
As the impedance of the environment is increased, the current-voltage 
characteristic (IVC) of the CPT develops a Coulomb blockade of Cooper pair tunneling and the
measured IVCs agree qualitatively with a theory based on quasicharge
dynamics for a CPT. Increasing the
impedance of the environment induces a transition in the modulation 
of the IVC with the gate charge from $e$-periodic to 2$e$-periodic.
\end{abstract}
\maketitle
\section{Introduction}

The interplay between charge quantization and superconductivity
offers the possibility to study macroscopic quantum phenomena in
small-capacitance Josephson junction circuits. One of the most
extensively studied devices in this context is the Cooper Pair
Transistor (CPT), which consists of two Josephson junctions in
series, with a gate capacitively coupled to the small island
formed between the two junctions.  Such single island circuits are
fundamental building blocks for quantum electronic circuits based
on the number-phase uncertainty of the Cooper pair condensate, and
have been the object of growing interest due to their potential to
realize new types of quantum limited measurements and solid state
quantum bits. Essential for all these applications is the ability
to inhibit the excitation of quasiparticles which give rise to
dissipation (noise), and thereby decoherence. For the CPT this
means that the charge parity of the island must be well
controlled, so that quasiparticles, or single unpaired electrons,
are not transferred to or from the CPT island.  Here we study the
CPT in a tunable high frequency electrodynamic environment, where
the DC current-voltage characteristic (IVC) of the CPT exhibits a
Coulomb blockade of Cooper pair tunneling.  By tuning the
impedance of the environment, we demonstrate how the parity of the
island is controlled by the environment.

CPTs have been most extensively studied in a high frequency
environment with low impedance, $Z_\mathrm{env}\ll R_Q\equiv
h/4e^2\simeq 6.45~\mbox{k}\Omega$, where the gate voltage
modulates the critical current flowing through the device.  At low
temperature and in the absence of quasiparticle excitations, this
modulation should be periodic in gate voltage with period
$2e/C_g$, where $C_g$ is the gate capacitance. $2e$-periodic
modulation is a clear sign that charge transport through the
device is due to Cooper pairs, which do not change the parity of
the island. However, in many experiments this parity effect is
spoiled by the transfer of quasiparticles to or from the island,
resulting in an $e$-periodic gate modulation, or modified
$2e$-periodic modulation, as the unpaired electron occupies the
island. For applications like the coherent control of a quantum
bit, these quasiparticles have to be avoided, and several
experimental studies have been carried out in order to understand
and control this parity effect. Normal metal leads, or small
normal metal ``traps", contacting the superconducting source and
drain close to the tunnel junctions are thought to help bring
long-lifetime out-of-equilibrium quasiparticles into thermal
equilibrium, and thereby enhance the parity
effect~\cite{joyez:parityeffect:94,agren:CPTtunable:2002}.
However, in many experiments where no traps were used, a parity effect was
demonstrated~\cite{geerligs:pairtunneling:90,tuominen:2eCPT:92,aumentado:2ePeriodCPT,
Amar:2eToeCPT:94}. Aumentado {\it et
al.}~\cite{aumentado:2ePeriodCPT} showed that the lifetime of the
unpaired quasiparticle on the island could be dramatically reduced
by fabricating the island of the CPT with a larger superconducting
energy gap than that of the source-drain.

Here we present an experimental study of the CPT in a high
impedance environment, $Z_\mathrm{env}\gg R_Q$, where far less
experimental work
has been done due to the difficulty in constructing such an environment. 
In a high impedance environment, the IVC of the  CPT shows not a supercurrent, 
but rather a Coulomb blockade
of Cooper pair tunneling. The Coulomb blockade is modulated with
the gate voltage and, in the absence of quasiparticles, this
modulation is $2e$-periodic. Previously, the high impedance
environment has been achieved with small on-chip resistors ($R\gg
R_Q$) located close to the CPT, which have enabled the observation
of a Coulomb blockade of Cooper pair tunneling with $2e$-periodic
dependence on the gate
charge~\cite{haviland:ssethighimpedance:94,zorin:SensitiveElec:99}.
More recently, one-dimensional SQUID arrays have been used to bias
a CPT~\cite{watanabe:SetsHighZEnv:04} which is the technique used
in the present study. The great advantage of the SQUID array bias
is that the impedance of the environment can be tuned {\it in
situ}, without affecting the CPT.  In this way the role of
fluctuations due to the environment can be easily identified.

In this article we study the IVC of the CPT as the
environment impedance is changed over many orders of magnitude.
Starting from a superconducting-like IVC at low impedance, a well
defined Coulomb blockade of Cooper pair tunneling with a
back-bending IVC develops as the impedance of the environment is
increased. This back-bending IVC is due to the coherent tunneling
of single Cooper pairs and is a clear indication that
$Z_\mathrm{env}\gg R_Q$. We show that the gate voltage dependence
of the IVC changes from $e$-periodic to $2e$-periodic as the
environment impedance is increased.  The high impedance
environment suppresses quasiparticle tunneling rates, thereby
restoring the even parity of the island. These observations
indicate new approaches for controlling out-of-equilibrium
quasiparticles in superconducting quantum circuits.

\section{Fabrication and Measurement}

\begin{figure}
\includegraphics{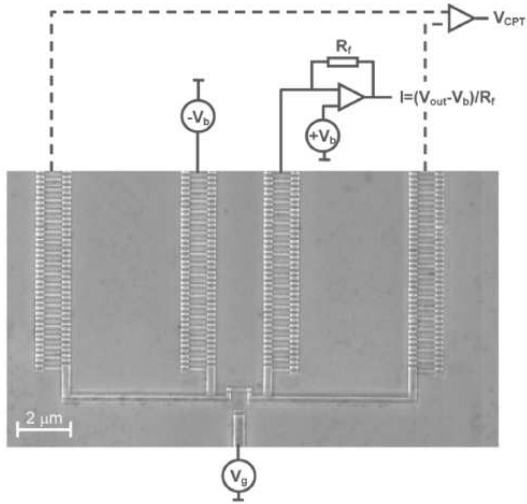}
\caption{\label{SEM}SEM micrograph of a CPT (in the center) biased
by four SQUID arrays. Two SQUID arrays are employed to apply a
symmetric bias, while the other pair is used to measure the
voltage across the CPT.}
\end{figure}

Figure~\ref{SEM} shows a SEM micrograph of a CPT sample biased by
four SQUID arrays. The Al/AlOx/Al tunnel junctions are fabricated
on a SiO$_{2}$/Si substrate using e-beam lithography and the standard
shadow evaporation technique~\cite{corlevi:thesis:06}. Four
arrays, each consisting of 70 SQUIDs in series, connect the CPT to
the bias and measurement circuitry. The SQUIDs are designed to
have two identical junctions in order to suppress the Josephson
coupling in the array as low as possible.  The nominal area of
each junction in a SQUID is 0.03~$\mu$m$^{2}$, and the loop area
is 0.18~$\mu$m$^{2}$. Each CPT junction has a nominal area of
0.01-0.02~$\mu$m$^{2}$. To provide an effective high impedance
environment, the arrays have to be located close to the CPT, so as
to avoid the shunting capacitance of the leads.

The SQUID arrays have high impedance due to the Josephson
inductance of the SQUID $L_{J}=\hbar/2eI_{C}$, which becomes
infinite as the critical current of the SQUID, $I_{C}$, is
suppressed to zero with an external magnetic flux. This inductance
is also non-linear, as it depends on the phase across the SQUID, and
therefore a linear circuit model can not give a complete
description of the electrodynamics of these arrays.  However, for
current much less than the critical current, a linear
approximation can be made, which shows that the infinite array
behaves like a transmission line with a gap, or region of
frequency where no propagating modes exist. A finite array will
have a set of discrete modes, and for arrays such as the ones 
used in this study, the lowest frequency mode
has $f\simeq 30$~GHz~\cite{haviland:1Dsum:2000,corlevi:thesis:06}.
The impedance of arrays such as these has not yet been
systematically characterized due to the very high frequencies
involved. However, we have shown that these arrays do provide an
environment which induces a Coulomb blockade of Cooper pair
tunneling and Bloch oscillations in a single Josephson
junction~\cite{watanabe:TuneEnviron:01,corlevi:DualityIZ:06}. As
the magnetic flux through the SQUID loops is increased, the
arrays undergo a quantum phase
transition~\cite{chow:SItrans1D:98,Kuo:SITrans1Darray:01} where
the measured zero bias resistance of the array increases several
orders of magnitude (50~k$\Omega < R_{0}< 1~$G$\Omega$).  A key
aspect of this type of experiments is that the arrays can be tuned
in this transition region to a point where they provide a high
enough impedance at high frequencies, while not having too large
$R_0$.  If $R_0$ becomes too large, the input offset current of
the voltage amplifier will inhibit the measurement of the
differential voltage across the CPT.

The chip with the arrays and CPT is bonded to a printed circuit
board with connectors, which is mounted in a RF-tight box on a
dilution refrigerator with a base temperature of 15~mK. The
cryostat leads are twisted constatan pairs, which act as rather
good microwave filters attenuating 35~dB/GHz up to 2~GHz, above
which a background level of -80~dB is measured with an open
cryostat. Additional room temperature RC filters were used to
suppress lower frequency noise when necessary.  No microwave filters
between the sample and twisted pairs were implemented in these measurements. However,
in the high impedance case, the arrays themselves are good filters,
protecting the CPT from electromagnetic fluctuations coming from
the bias side of the arrays.

The IVC of the CPT is measured in a four point configuration with
the CPT symmetrically biased through one pair of SQUID arrays,
while the other pair is used to measure the voltage across the
sample. The current is measured with a trans-impedance amplifier
(modified Stanford Research System 570) on one of the biasing
leads. The voltage across the CPT is measured with a high input
impedance differential voltage amplifier (Burr-Brown INA110). The
bias voltages $\pm V_b$ are applied symmetrically with respect to
ground in order to avoid a common mode drain-source voltage. In
this way, a polarization charge can be induced on the island only
by the voltage applied to the gate and not from a bias-induced
common mode voltage acting across the stray capacitance of the
CPT to ground \cite{corlevi:thesis:06}.

\begin{table*}
\caption{List of a few characteristic parameters of the measured
CPT samples.}
\begin{ruledtabular}
\begin{tabular}{cccccccc}
&$R_N$~(k$\Omega$)&$C$~(fF)&$E_{J}~(\mu$eV)&$E_{C}~(\mu$eV)&$E_{J}/E_{C}$&$V_{t}~(\mu$V)&$
V_\mathrm{gap}~(\mu$V)\\ \hline
CPT 6&5.2&0.75&125&106&1.2&15&35\\
CPT 7&21&0.6&30&133&0.22&40&110\\
CPT 8&1.5&0.8&430&100&4.3&5&/\\
CPT 9a&11.5&0.55&56&145&0.38&34&60\\
CPT 9b&8.2&0.55&80&145&0.55&26&45\\
\end{tabular}
\end{ruledtabular}
\end{table*}

The parameters of the measured samples are listed in Table~1. The
normal state resistance per junction $R_{N}$ in the CPT is calculated from
the slope of the IVC at large bias voltage, assuming that both junctions 
are identical. The Josephson energy is $E_{J}=(R_{Q}/R_{N})(\Delta/2)$, where
$\Delta\approx 200~\mu$eV is the superconducting energy gap of Al. The
capacitance of one junction in the CPT is calculated from the
junction area as measured from SEM pictures, using the specific
capacitance 45~fF/$\mu$m$^{2}$. From the capacitance $C$, the charging 
energy of one junction in the CPT is calculated as $E_{C}=e^2/2C$. 
For all the measured samples, $C_g \simeq 10~\mbox{aF}\ll C$, 
so that the total island capacitance is given
by $C_\Sigma =C_{1}+C_{2}+C_{g}\approx 2C$. In table~1, $V_{t}$
indicates the measured maximum threshold voltage for Cooper pairs,
as determined from the IVC of the CPT when $Z_\mathrm{env}\gg
R_{Q}$.
$V_\mathrm{gap}$ refers to the maximum value of the measured Coulomb gap for 
single electrons, estimated from the offset of the normal state IVC, 
taken in high magnetic fields
where superconductivity is completely suppressed.

\section{Current-voltage characteristic of the Cooper pair transistor in a high impedance environment}

\begin{figure}
\includegraphics[width=0.4\textwidth]{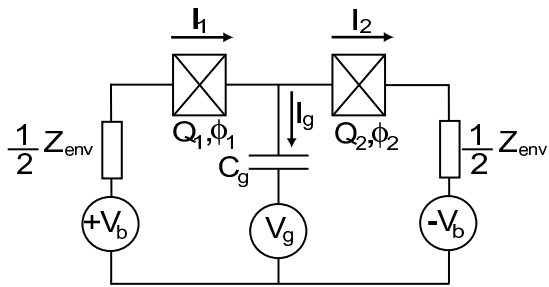}
\caption{\label{CPT_scheme} Schematic of a Cooper pair transistor.
The environment is represented by the impedance
$Z_\mathrm{env}$. When $Z_\mathrm{env}\ll R_{Q}$ ($Z_\mathrm{env}\gg R_{Q}$) 
the CPT is embedded in a low (high) impedance environment. $Q_{1(2)}$ indicates the charge and
$\phi_{1(2)}$ the phase difference across each junction.
$I_{1(2)}$ represents the current flowing through junction 1(2)
and $I_{g}$ the displacement current from the island to the gate
electrode.}
\end{figure}

Most experimental and theoretical studies to date have considered
the case of a CPT embedded in a low impedance environment,
$Z_\mathrm{env}\ll R_{Q}$~\cite{geerligs:pairtunneling:90,tuominen:2eCPT:92,
aumentado:2ePeriodCPT,Amar:2eToeCPT:94,joyez:parityeffect:94,
agren:CPTtunable:2002,matveev:parityeffect:93}. 
A few studies have been made for the case of a high impedance
environment, $Z_\mathrm{env}\gg R_{Q}$
~\cite{haviland:ssethighimpedance:94,zorin:SensitiveElec:99,watanabe:SetsHighZEnv:04}.
Below we review the theoretical description of the CPT for the low
impedance case, showing how it can be transformed to a description
appropriate for the high impedance case.  We then give a
qualitative description of the IVC and Bloch oscillations for the
later case, comparing this with experimental results.

We consider the circuit depicted in Fig.~\ref{CPT_scheme},
consisting of a symmetric CPT with $E_{J1}=E_{J2}=E_{J}$ and
$C_{1}=C_{2}=C$. The charge induced by the gate on the island is
$Q_{g}=C_{g}V_{g}$, where $V_{g}$ is the voltage applied to the
gate electrode. $Q_{1(2)}$ indicates the charge and $\phi_{1(2)}$
the phase difference across each junction. Let
$\phi=\phi_{1}+\phi_2$ be the bias phase across the CPT and
$\phi_{g}=(\phi_{1}-\phi_2)/2$ the phase of the island. The phases
$\phi$ and $\phi_{g}$ are conjugate respectively to the charge
transferred across the device $Q_{\phi}=(Q_{1}+Q_{2})/2$ and the
total island charge $Q=Q_{1}-Q_{2}+Q_{g}$ and obey the commutation
relations $[Q_{\phi},\phi]=[Q,\phi_{g}]=-2ie$.

For the moment we neglect the contribution of quasiparticles by
assuming that the superconducting energy gap $\Delta$ is large
compared to all relevant energy scales. The Hamiltonian of a CPT
with $C_{g}\ll C$ biased by a current $I_{b}=V_{b}/Z_\mathrm{env}$
can be written as follows

\begin{eqnarray}
H=8E_{C}\left(\frac{Q_{\phi}}{2e}\right)^2+2E_{C}\left(\frac{Q+Q_{g}}{2e}\right)^2
\nonumber\\
 -2E_{J}\mbox{cos}(\phi/2)\mbox{cos}(\phi_
{g}) -\frac{\hbar I_{b} \phi}{2e}.
  \label{HCPT1}
\end{eqnarray}

The first term of the Hamiltonian corresponds to the charging
energy associated with the total charge across the series
combination of the two junctions of the CPT. The second term is
the electrostatic energy associated with the island which sees the
two junctions in parallel. The third term is the effective
Josephson energy of the CPT, which, due to its dependence on
$\phi_{g}$, can be modulated by the gate voltage. The last term
represents the coupling between the bias current $I_{b}$ and the
phase $\phi$.

In a low impedance environment, $Z_\mathrm{env}\ll R_{Q}$, quantum
fluctuations of the phase are suppressed and the bias phase $\phi$
behaves as a classical variable. Due to the low impedance, Cooper
pair tunneling at zero bias voltage dominates the IVC of the CPT
which shows the typical superconducting behavior characterized by
a supercurrent branch at vanishing voltages. When $E_{J}\leq E_C$,
charging effects are observable and result in a $2e$-periodic
modulation of the critical current with the gate charge. Thus,
when $Z_\mathrm{env}\ll R_{Q}$, the CPT behaves as a single
Josephson junction with a gate charge dependent critical
current~\cite{joyez:parityeffect:94,agren:CPTtunable:2002}.

Let us turn to the case of a high impedance environment,
$Z_\mathrm{env}\gg R_{Q}$, where quantum fluctuations of the phase
are strong and the conjugate charge now becomes a classical
variable. For this case, a quantitative theory of the single
Josephson junction in terms of Bloch energy bands has been
developed~\cite{Averin:BlochOsc:85,likharev:blochoscillations:85}.
The dynamics of the junction can be described by the quasicharge
$q=\int{Idt}$, where $I$ is the current flowing through the
junction. The quasicharge evolves in time according to the
Langevin equation
\begin{equation}
\frac{dq}{dt}=(I_{b}+\delta I)-\frac{V(q)}{R}, \label{langeq}
\end{equation}
where $V(q)$ is the voltage across the junction and $\delta I$
indicates a random noise component of the bias current, induced by
the real impedance of the environment
$R=\mbox{Re}[Z_\mathrm{env}]\gg R_{Q}$ at finite temperature. The
voltage $V(q)$ is given by the derivative of the lowest Bloch
energy band, $V(q)=dE_{0}/dq$. The IVC of the junction can be
calculated analytically from the Langevin
equation~\cite{Averin:BlochOsc:85,likharev:blochoscillations:85,beloborodov:DualToIZ:02}
and shows a voltage peak near zero current corresponding to the
Coulomb blockade of Cooper pair tunneling, and a back-bending
region at finite currents. This region of negative differential
resistance is the indication of the Bloch oscillations, which
occur with frequency $f_{B}=\langle I\rangle /2e$.

\begin{figure}
\includegraphics{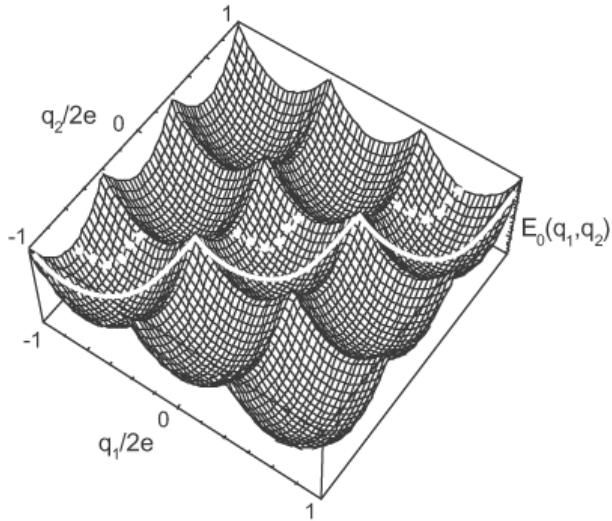}
\caption{\label{bands2D} Lowest energy Bloch band for a CPT with
$E_J/E_C\ll 1$. For fixed gate charges, the quasicharges $q_{1}$
and $q_{2}$ follow a diagonal trajectory. The solid and dotted
lines indicate respectively the quasicharges trajectories for
$Q_{g}=2ne$ and $Q_{g}=(2n+1)e$.}
\end{figure}

The natural extension of the Bloch band picture from a single
Josephson junction to a CPT leads to a description of the
quasicharge dynamics in two-dimensional Bloch energy bands. In
Fig.~\ref{bands2D} the lowest energy band of a CPT with
$E_J/E_C\ll 1$ is plotted in a space spanned by the two
quasicharges $q_{1}$ and $q_{2}$, which describe the dynamics of
the two junctions of the CPT. This description becomes clear if we
rewrite the Hamiltonian~(\ref{HCPT1}) in terms of the variables
$Q_{1(2)}$ and $\phi_{1(2)}$. For $I_{b}=0$ the
Hamiltonian~(\ref{HCPT1}) reads

\begin{eqnarray}
H=4E_{C}\left(\frac{Q_{1}}{2e}\right)^2&+&4E_{C}\left(\frac{Q_{2}}{2e}\right)^2+
\frac{2E_{C}}{e^2}(Q_{1}-Q_{2})Q_{g}
\nonumber\\&-&E_{J}\mbox{cos}(\phi_{1})-E_{J}\mbox{cos}(\phi_{2}).
 \label{HCPT2}
\end{eqnarray}

This Hamiltonian is that of two single Josephson junctions,
$H_{1(2)}=4E_{C}(Q_{1(2)}/2e)^{2} - E_{J}\mbox{cos}(\phi_{1(2)})$,
coupled by the term containing the gate charge $Q_{g}$. For small
coupling, $C_{g}\ll C$, the behavior of a CPT in a high impedance
environment is described by the quasicharges
$q_{1(2)}=\int{I_{1(2)}}dt$, where $I_{1(2)}$ indicates the
current flowing through junction 1(2) (Fig.~\ref{CPT_scheme}). In this case the
lowest energy band of the CPT is approximately given by the sum of
the lowest energy band of the two single junctions, that is
$E_{0}(q_{1},q_{2})\simeq E_{0}(q_{1})+E_{0}(q_{2})$.

The conservation of current results in the constitutive relation
$I_{1}= I_{2}+I_{g}$, where $I_{g}$ indicates the displacement
current flowing from the island of the CPT to the gate. From this
relation follows the condition $q_{1}-q_{2}=Q_{g}$, which, for a
fixed gate charge, constrains the quasicharges to a diagonal
trajectory in the two-dimensional Bloch band (Fig.~\ref{bands2D}).
A change in the gate voltage induces a shift in the trajectory of
the two quasicharges, as shown in Fig.~\ref{bands2D}. In analogy
with the one dimensional case, the voltage across the CPT can be
expressed via the derivative of the lowest energy band along the
diagonal direction as $V(q)=dE_{0}(q_{1},q_{2})/dq$, where
$q=(q_{1}+q_{2})/2$ defines the quasicharge of the CPT. The
maximum Coulomb blockade voltage of the
CPT, or critical voltage $V_{C}=\mbox{max}[dE_{0}(q_{1},q_{2})/dq]$, 
is a $2e$-periodic function of the gate charge. The dependence of the 
critical voltage on the gate charge is determined by the slope of the 
energy band according to the Bloch band theory~\cite{zorin:SensitiveElec:99}. 
The critical voltage is maximum when $Q_{g}=2ne$, where the slope 
of the lowest energy band is maximized (solid line in Fig.~\ref{bands2D}), 
while the minimum critical voltage is obtained for $Q_{g}=(2n+1)e$ 
(dotted line in Fig.~\ref{bands2D}).

Thus, when $Z_\mathrm{env}\gg R_{Q}$ the CPT behaves as a single
Josephson junction with a gate charge dependent critical voltage.
In the case $Q_g=2ne$, the energy-charge relation of the CPT is
identical to that of a single Josephson junction and the voltage
$V(q)$ across the CPT has a $2e$-periodic dependence on the
quasicharge. From the band diagram one can see that as the gate 
charge is shifted, the dependence of the voltage on the quasicharge
is modified. When $Q_{g}=(2n+1)e$, the Bloch oscillations in each
junction are out of phase, resulting in an oscillation
of the voltage across the CPT with the
quasicharge having frequency $f=e/\langle I \rangle$~\cite{zorin:SensitiveElec:99}. For $E_J>E_C$, the
lowest Bloch band becomes more sinusoidal along the quasicharge
axes, and the diagonal trajectory at $Q_g=(2n+1)e$ becomes flat
for a symmetric CPT.  In this case the Coulomb blockade for Cooper
pairs can be modulated to zero and the out of phase Bloch
oscillations cancel one another.

\begin{figure}
\includegraphics[width=0.4\textwidth]{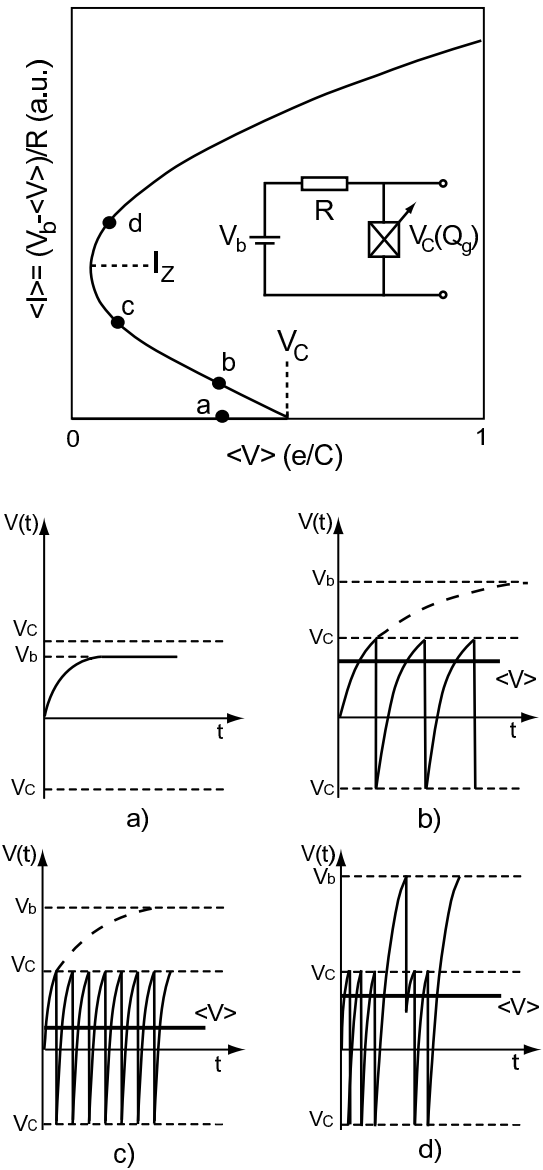}
\caption{\label{BO_CPT} A pictorial description of the Bloch
oscillations and back-bending IVC of a CPT, or single Josephson
junction with tunable critical voltage $V_C$, current biased by a
high impedance resistor. $I_{Z}$ indicates the Zener current. The
time dependence of the voltage across the junction is shown for 4
different operating points in the IVC. The shape of the Bloch
oscillations refers to the case of a junction with $E_J\ll E_C$.}
\end{figure}

The IVC of a CPT in the high impedance environment shows the same
features as that for a single Josephson junction and can be
calculated from the Langevin equation for the
quasicharge~(\ref{langeq}). We can picture how the Bloch
oscillations occur in the CPT, or single Josephson junction with
tunable critical voltage, by simply considering a non-linear
capacitor biased by a voltage source $V_b$  through a resistor
$R\gg R_{Q}$. In Fig.~\ref{BO_CPT} the generic shape of the IVC is shown,
with the time evolution of the junction voltage sketched for four
different operating points of the circuit.  We consider the case
$E_J<E_C$, where the non-linear capacitor has a periodic,
saw-tooth form of the voltage-charge relation, given by the
derivative of the lowest energy band along a diagonal trajectory
with a maximum critical voltage.
\begin{figure}
\includegraphics{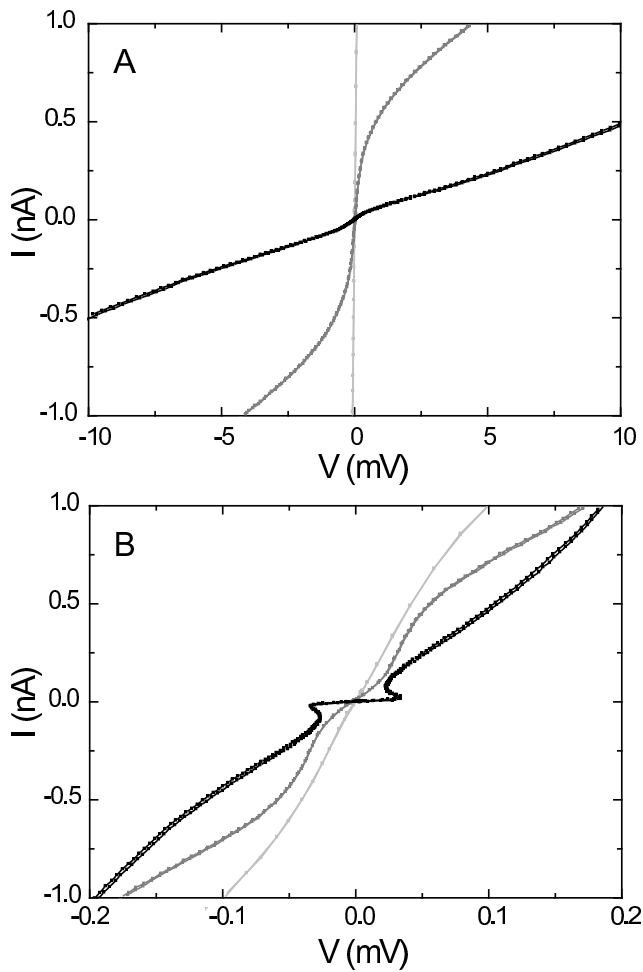}
\caption{\label{IVenv}(A): IVC of the two biasing SQUID arrays of
sample CPT~9a for different magnetic fields. The measured values
for $R_{0}$ are: Light-gray curve $R_{0}\approx$ 50~k$\Omega$,
dark-gray curve $R_{0}\approx $ 500~k$\Omega$, black curve
$R_{0}\approx $ 20~M$\Omega$. (B) Corresponding IVCs of CPT~9a.}
\end{figure}
When the bias voltage is below the critical voltage, $V_b<V_C$
(operating point a), the Langevin equation~(\ref{langeq}) will
have a stationary solution with $I=dq/dt=0$, and the non-liner
capacitor will simply charge up to the value $V_b$ with no voltage
oscillation.  As the bias is increased so that $V_b>V_C$, time
dependent solutions to the Langevin equation are possible, which
describe overdamped, forced Bloch oscillations in the junction.
These oscillations result in a periodic charging and discharging
of the non-linear capacitor, which lead to a time average voltage
$\langle V \rangle <V_C$ (operating point b). As the bias voltage
is further increased, the Bloch oscillations increase in frequency 
as higher current is forced through the junction.  With faster
Bloch oscillations, the finite charging time due to the bias
resistor becomes negligible, and the voltage oscillations become
more symmetric around zero, with $\langle V \rangle \rightarrow 0$
(operating point c).  Increasing the bias voltage further would
cause back-bending of the IVC to dissipation-less transport, or DC
current with zero average voltage.  However, with higher frequency
Bloch oscillations, Zener tunneling to higher energy bands takes
place, resulting in a temporary build-up of the junction voltage
to a level close to the superconducting energy
gap~\cite{schoen:review:90}.  In this case, quasiparticle
tunneling can no longer be neglected, and Josephson-quasiparticle
tunneling cycles, which are dissipative processes, can occur. As
the Zener tunneling probability increases with the bias current,
the time average voltage across the junction increases (operating
point d). The magnitude of the Zener current depends on the energy
gap between the lowest and first Bloch energy band, which in turn
depends on the ratio $E_J/E_C$~\cite{averin:singleelectronics:91,schoen:review:90}.

\begin{figure}
\includegraphics{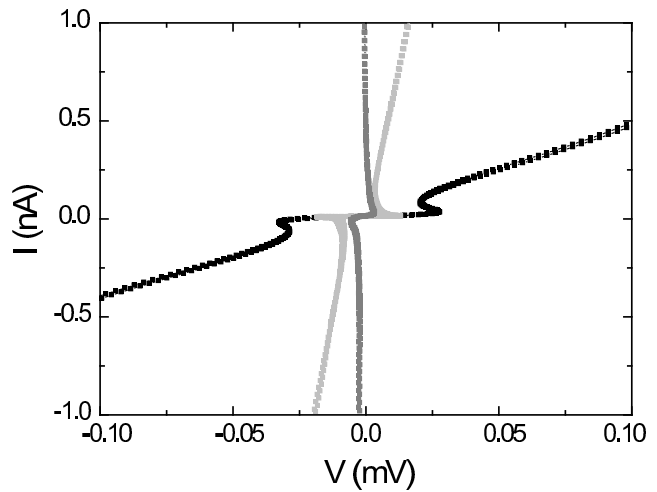}
\caption{\label{IVcpt} The IVC for three different CPTs having
approximately the same value of the zero bias resistance of the
SQUID arrays ($R_0\approx 20~$M$\Omega$). Black curve: CPT 9a,
$E_J/E_C =0.38$, Light-gray curve: CPT 6, $E_J/E_C=1.2$ Dark-gray
curve: CPT 8, $E_J/E_C =4.3$.}
\end{figure}

We find qualitative agreement with this theoretical picture in our
experimental IVCs of CPTs biased by SQUID arrays.
Figure~\ref{IVenv}A shows the IVC of two biasing SQUID of sample
CPT~9a measured in series at three values of the magnetic field,
between zero and $\Phi_0 / 2$. Here we see that the arrays change
their zero bias resistance $R_0$ over several orders of magnitude
as the Josephson coupling of the SQUIDs is suppressed.  In
Fig.~\ref{IVenv}B the corresponding IVCs of the CPT are plotted
for the same magnetic fields.  At low $R_0$ of the arrays, the IVC
of the CPT shows a superconducting-like feature that disappears as
the Coulomb blockade region becomes progressively more defined for
increasing values of $R_0$. When $R_{0}\approx 20$~M$\Omega$, the
Coulomb blockade for Cooper pairs is fully developed and is
followed by a region of negative differential resistance. This
back-bending IVC is the hall mark feature of a Coulomb blockade of
Cooper pair tunneling, and clear evidence that a high impedance of
the environment has been achieved. Figure~\ref{IVcpt} shows the
IVC of three different CPT samples with different values of
$E_J/E_C$.  In each case the SQUID arrays are tuned to
approximately the same value of $R_0$ where the back-bending IVC
is well defined. Figure~\ref{IVcpt} clearly demonstrates
qualitative agreement with the theory described above. With
increasing $E_J/E_C$ ratio, the critical voltage decreases, as the
lowest Bloch band develops more shallow minima.  We also see that
the Zener current, becomes larger with increasing $E_J/E_C$
ratio, as the gap to higher energy bands becomes larger.

\section{Gate voltage modulation and impedance dependent parity effect}

\begin{figure}
\includegraphics{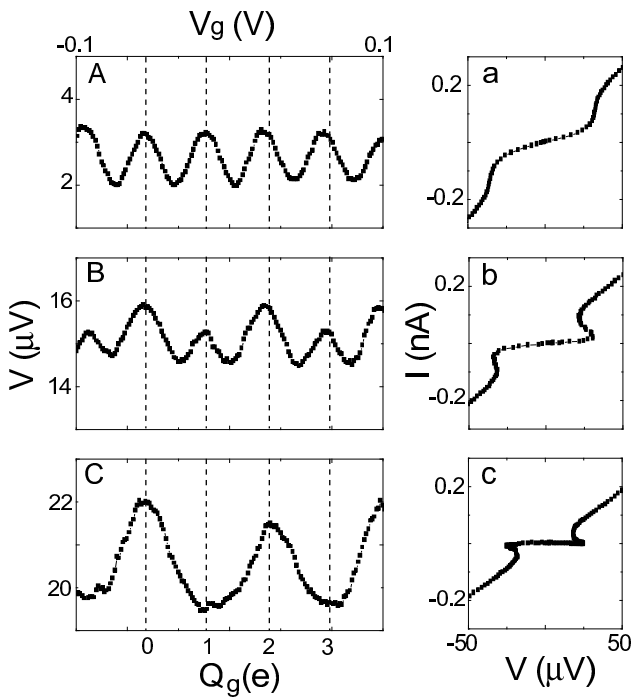}
\caption{\label{e2e_env}(A,B,C): Gate-induced voltage modulation
of sample CPT~9a for increasing values of the zero bias resistance
of the biasing SQUID arrays. The bias current is fixed at $I_{b}=
5~$pA. (a,b,c): IVC of the CPT for the same values of $R_{0}$.
(A,a): $R_{0}\approx 1$~M$\Omega$, (B,b): $R_{0}\approx
5$~M$\Omega$, (C,c): $R_{0}\approx 40$~M$\Omega$.}
\end{figure}

As discussed in the previous section, the two-dimensional Bloch
band picture describing transport in the CPT in terms of the
quasicharge leads to a 2$e$-periodic modulation of the IVC with
gate voltage.  This picture only considers Cooper pair tunneling,
completely neglecting quasiparticle or single electron tunneling.
If the single electron tunneling rates are large enough, they can
disrupt this picture, causing $e$-periodic modulation of the IVC.
In our experiments we find that the impedance of the environment
can be used to suppress these quasiparticle tunneling rates low
enough for stable 2$e$-periodic behavior to be observed.
Figure~\ref{e2e_env} shows how the periodic modulation of the
threshold voltage with gate at a constant current $I_{b}=5~$pA
evolves from 2$e$-periodic to $e$-periodic as the impedance of the
environment is tuned.  At high impedance, where the Coulomb
blockade is well developed, we find 2$e$-periodic behavior, as
expected from the two-dimensional Bloch band picture
(Fig.~\ref{e2e_env}C).  As the impedance of the environment is
reduced, the effect of ``quasiparticle poisoning" is seen as a
small peak in the threshold voltage at gate charges $Q_g =
(2n+1)e$ (Fig.~\ref{e2e_env}B), which grows until an $e$-periodic
behavior is observed (Fig.~\ref{e2e_env}A). A similar crossover
from 2$e$ to $e$-periodic behavior as the environment impedance is
tuned has been observed by Watanabe~\cite{watanabe:SetsHighZEnv:04}
and Kuo {\it et. al.} \cite{Kuo:posterMSS2006:06}.

To understand how quasiparticles affect the critical voltage of a
symmetric CPT in a tunable environment it is instructive to
determine the regions in the bias voltage -- gate voltage plane
where it is energetically favorable for a single quasiparticle to
tunnel on or off the island, thereby changing the island parity.
As we will see, the shape of the corresponding stability diagrams
depends on the impedance of the environment as well as on the
ratio of the two characteristic energies for adding or removing a
single quasiparticle to or from the island. These energies are the
charging energy $E_{C_{\Sigma}} = e^{2}/2C_\Sigma$ of the island
and the free energy difference $F=\Delta - k_B T
\mbox{ln}(N_{eff})$ that lifts the odd parity from the even parity
states~\cite{tuominen:2eCPT:92,joyez:parityeffect:94,matveev:parityeffect:93}.  
Here, the entropy term
involves $N_{eff} = N(0) V \sqrt{2 \pi \Delta T}$, the number of
quasiparticle states available within temperature $T$ just above
the gap ($N(0)$ is the normal-state density of states at the Fermi
level and $V$ the volume of the island). At zero temperature,
$N_{eff} = 0$ and this energy difference is the superconducting
energy gap $\Delta$, which is paid for having one unpaired
electron on the island. The free energy difference $F$ vanishes at
a temperature $k_B T^* = \Delta / \mbox{ln}(N_{eff})$, generally
much lower than the superconducting gap.

Let us first consider the situation of a free energy difference 
smaller than the island charging energy, $F<E_{C_{\Sigma}}$.
In Fig.~8a, we plot the corresponding stability diagram for the
case of a low impedance environment, where tunneling is elastic
and we can apply the global rule~\cite{ingold:rates:92}. The
stability diagram contains lines which separate regions in the bias
voltage -- gate voltage plane with stable parity from regions
where it is energetically favorable for a single quasiparticle to
tunnel on or off the island, thereby changing the parity. Within
the dotted rhombi the transition rate $\Gamma _\mathrm{oe}$ from
odd to even parity is zero, while within the solid rhombi the rate
from even to odd parity $\Gamma _\mathrm{eo}$ vanishes. Outside
the rhombi, a parity change corresponds to a decrease in energy
and the transition rates $\Gamma _\mathrm{oe}$ and $\Gamma
_\mathrm{eo}$ are non-zero at $T=0$. We conclude that the odd parity states
are stable near the gate charges $Q_g = (2n+1)e$ (Fig.~8a). At these gate
charges the system will be ''poisoned" by the excess quasiparticle
which sits on the island in the lower energy state. When in the
odd parity state, the effective island charge is shifted by $e$
and we thus expect the gate voltage dependence of the critical
voltage of the CPT to be shifted correspondingly (from the solid
to the dashed line in Fig.~8c). In particular, as gate voltage is
changed, the critical voltage should have two maxima per
$2e$-period (near gate charges $Q_g = (2n+1)e$ and $Q_g =2ne$).
This is in qualitative agreement with the low impedance data of
Fig.~7A and 7B.

In a high impedance environment, quasiparticle tunneling is
inelastic as a tunnel event is accompanied by energy exchange with
the environment and the local rule~\cite{ingold:rates:92}
should be applied to establish the stability diagrams. As a result, the size
of the rhombi doubles, and various stability regions start to
overlap. Overlap means that there are bistable regions where either
parity is stable, depending on the history of system. However as can
be seen in Fig.~8b, for $F<E_{C_{\Sigma}}$, odd stability regions
(dashed rhombi) still exist near gate charges $Q_g = (2n+1)e$, and
we expect the critical voltage to show two maxima per $2e$ period.
This observation is in contradiction with the experimental high
impedance data shown in Fig. 7C, where only one critical voltage
maximum per $2e$ period is observed, near gate charges $Q_g = 2ne$.

We therefore consider the stability diagram for the case $F \sim
E_{C_{\Sigma}}$, first in a low impedance environment, as shown in
Fig.~8d. Compared to the case $F < E_{C_{\Sigma}}$ (Fig.~8a), the
odd stability regions now disappear, giving rise to regions near
the gate charges $Q_g = (2n+1)e$  where both rates $\Gamma
_\mathrm{oe}, \Gamma _\mathrm{eo} \ne 0$ and parity will
fluctuate at small bias voltage.  However, in a high impedance 
environment for $F \sim E_{C_{\Sigma}}$, as shown in Fig.~8e,  
we observe that the odd stability regions completely
disappear, giving way to overlapping regions of even stability
only, as long as the applied bias voltage satisfies the
inequality $V_\mathrm{b} < e/C - 2F/e $. This is markedly
different from the low-impedance case, where regions of fluctuating
parity exists at any finite bias. We thus conclude that the
critical voltage is $2e$-periodic as a function of gate voltage (Fig.~8f) for
high impedance and $F\sim E_{C_{\Sigma}}$, in agreement
with the experimental findings of Fig.~7C.

This stability analysis shows that quasiparticle poisoning can occur near
the gate charges $Q_g = (2n+1)e$ if conditions are not favorable.  
The measured, long-time average CPT voltage 
near these gate charges will depend on the relative weight 
of the odd and even states, which are fluctuating
on a much more rapid time scale determined by the rates
$\Gamma _\mathrm{oe}, \Gamma _\mathrm{eo}$.  The value of these
rates cannot be inferred from the stability analysis
alone~\cite{schoen:parityNSN:94}.  For $\Gamma _\mathrm{oe} \gg \Gamma _\mathrm{eo}$
the effective island charge will most often be
even, hence the measured voltage of the CPT will be minimal near
$Q_g = (2n+1)e$ (see Fig.~8f). The effects of poisoning will not
be noticed on the long time scale of the experiment and one
observes a stable 2$e$-periodic modulation of transport. In the
opposite case $\Gamma _\mathrm{oe} \ll \Gamma _\mathrm{eo}$ near
$Q_g = (2n+1)e$, the decay of the odd state is much slower than
that of the even state. The effective island charge will be
predominantly odd and the measured average
voltage, instead of taking a minimum value, will reach a local
maximum, that eventually may become equal to the maxima found at
$Q_g = (2n)e$ (see Fig.~8c). The measurements for low-impedance 
presented in Fig.~7A show $e$-periodicity of the
critical voltage with gate voltage, we conclude that the ratio
$\Gamma_\mathrm{oe} \ll \Gamma_\mathrm{eo}$. This is in agreement
with several other experiments on CPTs with $F \sim
E_{C_{\Sigma}}$ in low impedance environments
~\cite{agren:thesis:02,aumentado:2ePeriodCPT,Wal:thesis:2001,eiles:josephsoncharging:94}.
We also note that the observed $e$-periodicity excludes that
$F$ is large compared $E_{C_{\Sigma}}$. If $F> E_{C_{\Sigma}}$, we
would have found overlapping regions of stable even parity near
$Q_g = (2n+1)e$, rather than regions of fluctuating parity and the
resulting dependence of critical voltage on gate voltage would
have been $2e$-periodic in the low impedance case.

We finally comment on the effect of finite temperature. The energy
difference $F$ depends on temperature $T$. For the islands used in
this experiment, the characteristic crossover temperature $T^*$
where $F \rightarrow 0$ can be estimated to be approximately
250~mK. The temperature $T$ appearing in the expression for $F$
does not necessarily correspond to the cryostat temperature, and
could be an effective temperature, describing the fluctuations in
the environment. Previous analysis of single junctions IVCs biased
by similar arrays~\cite{corlevi:DualityIZ:06} showed an effective
noise temperature approximately 150~mK, which would result in $F
\sim E_{C_{\Sigma}}$ in agreement with our conclusions above. A
second effect of thermal fluctuations is that they mix states of
different parity whose energy difference is $\alt k_B T$. On the
level of the stability diagrams this means that the lines
separating stability regions of different parity smear into strips
whose width is $\sim C k_B T$. For the low impedance case this has
drastic effects, even at low bias, because it leads to broadening
of the parity fluctuation regions near $Q_g = (2n+1)e$ thereby
enhancing the $e$-periodic gate modulation. For high impedance the
even parity stability region has a gap with respect to the bias
voltage, therefore thermal fluctuations do not effect the
$2e$-periodicity as long as $k_BT <e^{2}/C - 2F \sim e^{2}/2C$.
The fact that parity stability is robust against fluctuations for
small bias voltage in a high impedance environment is directly
connected with the inelastic nature of tunneling in this limit.
This inelasticity is at the origin of the bias voltage gap and
leads to an exponential suppression of quasiparticle tunneling at
low bias voltage. This effect is absent for low impedance as
tunneling is elastic.

\begin{figure}
\includegraphics[width=0.48\textwidth]{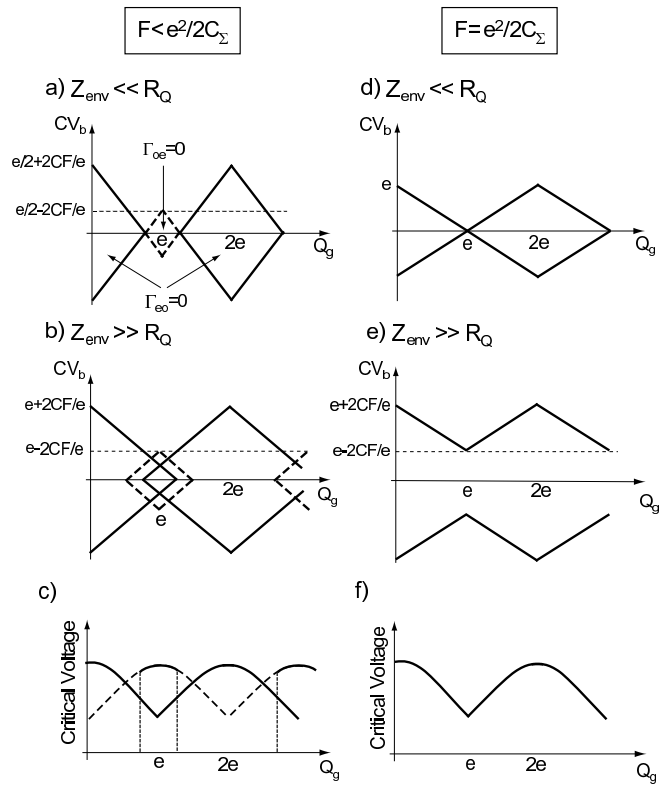}
\caption{\label{poison}a,b,c,d: Stability diagrams for the cases
$F < E_{C_{\Sigma}}$ with $Z_\mathrm{env}\ll R_{Q}$ and
$Z_\mathrm{env}\gg R_{Q}$ (a and b), $F = E_{C_{\Sigma}}$ with
$Z_\mathrm{env}\ll R_{Q}$ and $Z_\mathrm{env}\gg R_{Q}$ (d and e).
c) $e$-periodic gate voltage dependence of the critical voltage at
finite bias voltage ($V_{b}<e/C-2F/e$) corresponding to the
stability diagrams of a,b and d. f) $2e$-periodic gate voltage
dependence of the critical voltage at finite bias voltage
($V_{b}<e/C-2F/e$) corresponding to the case e).}
\end{figure}

\begin{figure}
\includegraphics{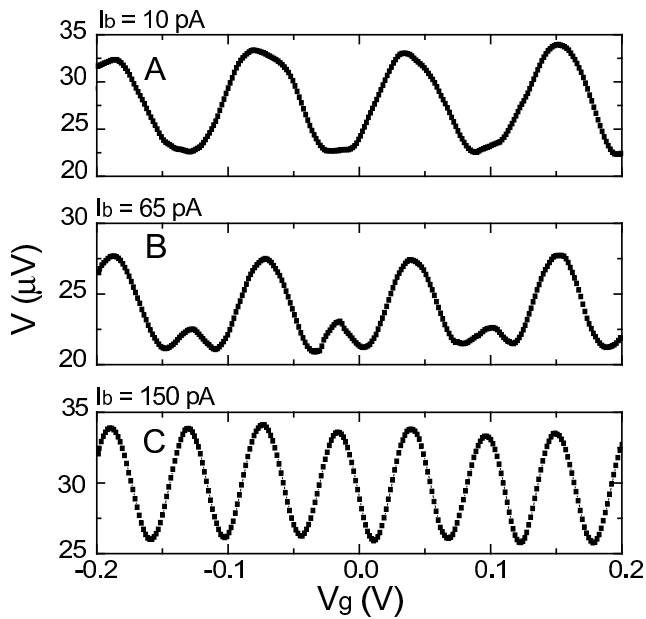}
\caption{\label{e2ebias} Gate voltage modulation for sample CPT~9a
at various bias current. The zero bias resistance of the biasing
SQUID arrays is $R_{0}\approx 20$~M$\Omega$. The IVC of the CPT
for this magnetic field is shown in Fig.~\ref{IVenv}.}
\end{figure}

The condition for stable $2e$-periodic behavior is not preserved at all
bias currents in the high impedance environment.  Figure~\ref{e2ebias}
shows the gate voltage modulation for high impedance
($R_0\approx 20$ M$\Omega$) of the CPT voltage for three different values of the bias current.
>From low current levels up to the back-bending region of the IVC,
the voltage modulation is stable 2$e$-periodic
(Fig.~\ref{e2ebias}A). As the current is increased toward the
Zener current, which for this sample is measured at $I_{Z}\approx
75$~pA, the rate of single electron transitions increases, as
explained in the previous section.  Zener tunneling causes an
increase in the rate $\Gamma _\mathrm{eo}$ near $Q_g = (2n+1)e$,
and the effect of poisoning can be observed (Fig.~\ref{e2ebias}B).
At even higher currents, the quasiparticle tunneling rate is so
large that only $e$-periodic modulation can be seen
(Fig.~\ref{e2ebias}C).

\section{Conclusions}

The effect of the electrodynamic environment on single charge
tunneling in the Cooper Pair Transistor (CPT) has been studied
experimentally by biasing and measuring the CPT through
one-dimensional SQUID arrays.  The impedance of these arrays can
be tuned with an external magnetic field without changing the
parameters of the CPT, which allows one to clearly discern the
effect of the environment on tunneling.  As the impedance of the
arrays is increased, the CPT develops a Coulomb blockade of Cooper
pair tunneling with a well defined back-bending current-voltage
characteristic (IVC), indicative of the coherent transfer of
single Cooper pairs. The general shape of the measured IVC is that
predicted by a theory based on quasicharge dynamics. We find that
the observed threshold voltage depends on the ratio between the
Josephson energy and the charging energy in a way that is in
qualitative agreement with the Bloch band theory for the CPT. As
the impedance of the environment is increased, we observe a
transition from $e$ periodic to $2e$-periodic modulation of the
IVC, demonstrating that the high impedance environment restores the
parity effect in the CPT. Enhanced parity effect, or reduced quasiparticle poisoning,
is explained by the inelastic nature of single-charge tunneling in a high
impedance environment, where a charging energy gap makes odd parity states
unstable near gate charges $Q_{g}=(2n+1)e$.

\section{Acknowledgments}

This work was partially supported by the Swedish VR, SSF
NanoDev program and the EU project SQUBIT. Samples were made in
the KTH Nanofabrication lab, provided by the K. A. Wallenberg
foundation. FWJH acknowledges support from the Institut Universitaire de France'.


\end{document}